\documentclass[onecolumn,12pt]{article}
\usepackage{amsmath}
\usepackage{booktabs}  
\usepackage{graphicx}
\usepackage{mathrsfs}
\usepackage{amssymb}
\usepackage{color}

\usepackage{url}

\usepackage{setspace}
\usepackage{authblk}
\usepackage{hyperref}
\hypersetup{
    colorlinks=true,    
    linkcolor=blue,     
    citecolor=blue,     
    urlcolor=blue       
}

\usepackage{multirow}
\usepackage{makecell}
\usepackage{caption}
\usepackage{caption}
\usepackage[backend=biber,
            style=nature,
            sorting=none,
            doi=false,
            url=true,      
            eprint=false,
            isbn=false,
            date=year,
            autocite=superscript]{biblatex}

\AtEveryBibitem{\clearfield{month}}
\AtEveryCitekey{\clearfield{month}}

\AtEveryBibitem{\clearfield{issn}}
\AtEveryCitekey{\clearfield{issn}}
\AtEveryBibitem{\clearfield{issue}}
\AtEveryCitekey{\clearfield{issue}}

\DeclareFieldFormat{journal}{#1} 
\DeclareFieldFormat{volume}{#1} 
\DeclareFieldFormat{pages}{#1} 
\DeclareFieldFormat{year}{#1} 
\DeclareFieldFormat{article}{%
  \href{https://doi.org/\thefield{doi}}{%
    \mkbibemph{#1}, \textbf{\printfield{volume}} (\printfield{year}), \printfield{pages}%
  }%
}
\defbibfilter{notinsegment1}{not segment=1}

\usepackage{lineno}
\usepackage{titlesec}  

\renewcommand{\thesection}{\Alph{section}}         
\renewcommand{\thesubsection}{\thesection.\arabic{subsection}} 
\titleformat{\section}
  {\large\bfseries}    
  {\thesection}        
  {1em}                
  {}                   

\titleformat{\subsection}
  {\normalsize\bfseries} 
  {\thesubsection}       
  {1em}                  
  {}                     
  
\usepackage{geometry} 
\makeatletter
\renewcommand{\@maketitle}{%
  \newpage
  \vspace*{2em} 
  {\large\bfseries\centering \@title \par} 
  \vspace{0.5em} 
  {\normalsize\centering 
    \lineskip .5em%
    \begin{tabular}[t]{c}%
      \@author 
    \end{tabular}\par}
  \vspace{0.5em} 
  {\normalsize\centering \@date} 
}
\makeatother

\title{Rare-Event-Induced Ergodicity Breaking in Logarithmic Aging Systems}
\author[1, 2]{Chunyan Li}
\author[1]{Qingyang Feng}
\author[1,2]{Tianjie Zhou}
\author[1,2,3]{Haiwen Liu\textsuperscript{\(\star\)}}
\author[3,4]{X. C. Xie}
\affil[1]{\textit{Center for Advanced Quantum Studies, Department of Physics, Beijing Normal University, Beijing 100875}}
\affil[2]{\textit{Key Laboratory of Multiscale Spin Physics, Ministry of Education, Beijing 100875, China}}
\affil[3]{\textit{Interdisciplinary Center for Theoretical Physics and Information Sciences, Fudan University, Shanghai 200433, China}}
\affil[4]{\textit{International Center for Quantum Materials, School of Physics, Peking University, Beijing 100871, China}}
\date{}

\begin{document}
\onehalfspacing

\maketitle
\noindent {\textsuperscript{*}Corresponding author. E-mail: haiwen.liu@bnu.edu.cn (H.L.)}
\begin{refsegment}
\begin{abstract}
Ergodicity breaking\cite{ doi:10.1103/PhysRev.109.1492, doi:10.1103/PhysRevLett.43.1754,
https://doi.org/10.1016/j.aop.2005.11.014,doi:10.1103/PhysRevB.75.155111} and aging effects\cite{refId0,doi:10.1103/PhysRevLett.71.173,DOI10.1088/0305-4470/29/14/012,doi:10.1142/9789812819437_0006,10.1007/978-3-540-44835-8_7, 10.1103/RevModPhys.83.587} are fundamental challenges in out-of-equilibrium systems. Various mechanisms\cite{refId0,doi:10.1103/PhysRevLett.71.173,10.1103/RevModPhys.83.587,PhysRevLett.87.055502,doi:10.1073/pnas.1003693107} have been proposed to understand the non-ergodic and aging  phenomena, possibly related to observations in systems ranging from structural glass\cite{PhysRevLett.75.2847} and Anderson glasses\cite{Ovadyahu_PhysRevLett.92.066801} to biological systems\cite{13prl_biology_PhysRevLett.88.158101,13prl_biology_nature} and mechanical systems\cite{log_aging_nature,PhysRevLett.118.085501}. While anomalous diffusion\cite{Shlesinger1993,Montroll1984,BOUCHAUD1990127} described by Lévy statistics\cite{10.1093/oso/9780198537885.003.0004} efficiently captures ergodicity breaking\cite{PhysRevLett.101.058101}, the origin of aging and ergodicity breaking in systems with ultraslow dynamics remain unclear. Here, we report a novel mechanism of ergodicity breaking in systems exhibiting log-aging diffusion. This mechanism, characterized by increasingly infrequent rare events with aging, yields statistics deviating significantly from Lévy distribution, breaking ergodicity as shown by unequal time- and ensemble-averaged mean squared displacements and two distinct asymptotic probability distribution function. Notably, although these rare events contribute negligibly to statistical averages, they dramatically change the system's characteristic time. This work lay the groundwork for microscopic understanding of out-of-equilibrium systems and provide new perspectives on glasses and Griffiths–McCoy singularities.
\end{abstract}

\maketitle

\subsection*{Introduction}

Ergodicity, a cornerstone of equilibrium statistical physics, is typically defined as the equivalence between the ensemble average and the time average of observables in the long-time limit\cite{Sethna2021}. Nevertheless, ergodicity breaking is widespread in out-of-equilibrium and non-stationary systems, arising from various mechanisms\cite{doi:10.1142/9789812819437_0006,10.1007/978-3-540-44835-8_7}. For example, in spin glass systems\cite{doi:10.1103/PhysRevLett.71.173} ergodicity breaking is primarily driven by the infinitely long time required for the system to explore all possible states. In addition, ergodicity breaking may result from the effect of Anderson localization\cite{ https://www.nature.com/articles/nature07000, https://www.nature.com/articles/nature07071}, many-body localization\cite{ https://www.science.org/doi/full/10.1126/science.aaa7432} and integrable Bose gases\cite{Kinoshita2006}. Non-ergodic behavior is also observed in biological systems\cite{PhysRevX.12.031005, PhysRevLett.106.048103}.

Aging, a key characteristic in out-of-equilibrium systems, refers to the phenomenon in which the relaxation dynamics slow progressively over time, exhibiting ultraslow behavior\cite{doi:10.1142/9789812819437_0006,10.1007/978-3-540-44835-8_7, 10.1103/RevModPhys.83.587}. This effect, where the system’s response depends on the elapsed time since its initialization, has been experimentally observed in structural glass\cite{PhysRevLett.87.055502}, Anderson glass\cite{Ovadyahu_PhysRevLett.92.066801}, crumpled thin sheets\cite{PhysRevLett.118.085501}, and disordered systems\cite{log_aging_nature}. Moreover, Bernaschi et al.\cite{doi:10.1073/pnas.1910936117} demonstrates ergodicity breaking in the aging process of spin glasses on a complex energy landscape. A central challenge remains in characterizing the complex dynamics of out-of-equilibrium systems, particularly those exhibiting both aging and ergodicity breaking.

Log-aging diffusion provides a unique testing ground to re-examine the mechanisms of ergodicity breaking, aging and the subtle role of rare events beyond conventional statistical descriptions. Anomalous diffusion, observed in various complex systems\cite{PhysRevB.12.2455, doi:10.1073/pnas.1016325108}, is characterized by the mean squared displacement (MSD) growing asymptotically as a power law, \(\langle \Delta^2 x \rangle \propto t^\alpha\) (with \(\alpha = 1\) corresponding to normal diffusion).  Previous studies\cite{PhysRevLett.90.120601, PhysRevLett.93.190602} suggest that ergodicity breaking is related to the occurrence of L\'evy statistics\cite{Shlesinger1993}, and dominated by a few rare events\cite{PhysRevB.66.052301, RevModPhys.62.251}. In this context, theories have been proposed to explain out-of-equilibrium dynamics\cite{doi:10.1073/pnas.1003693107,doi:10.1073/pnas.1120147109, D2CP01741E}. However, the origin of aging and its influence on ergodicity remain unclear, especially in systems with log-aging diffusion characterized by \(\langle \Delta^2 x \rangle \propto \ln{(t/t_0)}\) with \(t_0\) representing the initial time\cite{PhysRevLett.110.208301, li2024routerandomprocessultraslow,Sanders_2014,PhysRevX.15.011043}, related to various physical systems\cite{Ovadyahu_PhysRevLett.92.066801,13prl_biology_nature,13prl_biology_PhysRevLett.88.158101,log_aging_nature}. 

Through analysis of log-aging diffusion, we demonstrate ergodicity breaking governed by an unconventional mechanism. We find that increasingly infrequent rare events paradoxically control the long-time dynamics, resulting in long-tail distributions and infinite characteristic time. These findings pertain to ultraslow dynamics modeled via rugged landscapes\cite{refId0} and demonstrate dynamics analogous to Griffiths-McCoy singularities\cite{PhysRevB.51.6411}.

{\subsection*{Scale invariance breaking due to aging}}

\begin{figure*}[hbt] 
\centering
\includegraphics[width=1\textwidth]{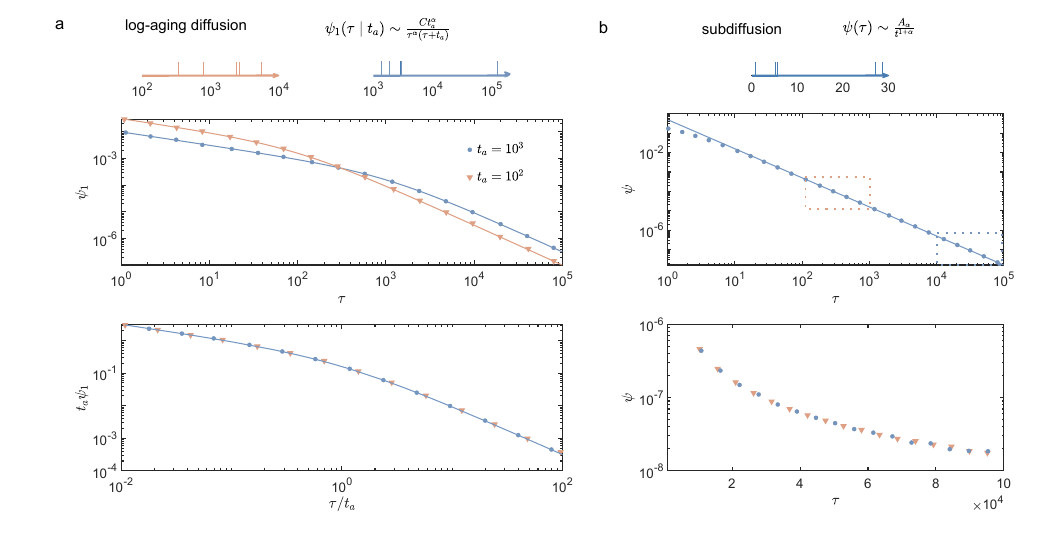} 
\caption{ \textbf{Scheme of Log-Aging Diffusion and Subdiffusion}.  
(a) The forward waiting time distribution of log-aging diffusion governs the dynamics of the process, thereby introducing aging effects. Simulations of the first \(5\) steps' random waiting times with \(\alpha=0.5\) for \(t_0=10^2\) and \(10^3\) are displayed in the top panel. In the bottom panel, the forward waiting time distributions at different \(t_a\) values collapse into a single curve after scaling by the factor \(t_a\). (b) The waiting time distribution of subdiffusion exhibits a power-law tail, ensuring scaling invariance. In the top panel, the waiting times are distributed according to \(\frac{\alpha}{(t+1)^{1+\alpha}}\), \(\alpha=0.5\). As shown in the bottom panel, data in the interval \(10^2-10^3\), when scaled by a factor of \(10^{2}\) (solid triangles), reproduce the features observed in the interval \(10^4-10^5\) (blue dots).}
\label{fig1}
\end{figure*}
The log-aging diffusion\cite{PhysRevLett.110.208301, li2024routerandomprocessultraslow}, based on continuous-time random walk (CTRW)\cite{10.1063/1.1704269, Montroll1984, RalfMetzler_2004}, offers a statistical framework to analyze complex systems. In log-aging diffusion, both the jump length and the waiting time between successive jumps are continuous and random. While the jump length distribution is trivial, the waiting time distribution is flexible and depends on the present time \(t_a\), also referred to as the forward waiting time distribution\cite{PhysRevLett.110.208301}, and is expressed as \(\psi_1(\tau\mid t_a) = \frac{\sin{\pi\alpha}}{\pi} \frac{t_a^\alpha}{\tau^\alpha(\tau+t_a)}\). One consequence of this is that the process is sensitive to the initial time, \(t_0\), of the experiment, such as the time when the external field is withdrawn. The logarithmic time evolution of log-aging diffusion for initial times \(t_0=10^2\) and \(10^3\) is presented, and the statistics of waiting times at different present times \(t_a\) are shown in Fig~\ref{fig1}a. The system undergoes aging: over time, its state continues to depend on the current time, making it more likely to experience longer waiting periods and thus slowing down progressively. Moreover, the forward waiting time distribution decays with a power-law tail, leading to an infinite average waiting time. The collapse of the curves for different \(t_a\) when plotted with \(t/t_a\) on the x-axis and \(t_a\psi_1\) on the y-axis indicates that the system's scale varies over time, thereby breaking its scale invariance. These features make log-aging diffusion a powerful tool for analyzing the dynamics of out-of-equilibrium disordered systems, which are characterized by a nontrivial energy landscape and an infinite phase space of lifetime quantities\cite{refId0}.

For subdiffusion, often modeled as a renewal process, the waiting time distribution \(\psi(\tau)\) exhibits an asymptotic power-law tail, \(\psi(\tau) \sim A_\alpha \tau^{-1-\alpha}\). This heavy tail leads to time-independent scale invariance in the system's properties (Supplementary Fig.~\ref{supplement_fig.sub}). As demonstrated by data collapsing across different time scales (Fig.~\ref{fig1}b), this scale invariance governs key characteristics, such as constant relative fluctuations\cite{PhysRevLett.101.058101} and specific scaling forms for probability distributions\cite{PhysRevLett.130.207104}. Thus, the waiting time distribution fundamentally determines the system's characteristic relaxation dynamics.\\

{\subsection*{Ergodicity breaking in logarithmic aging systems}}

\begin{figure*}[hbt] 
\centering
\includegraphics[width=1\textwidth]{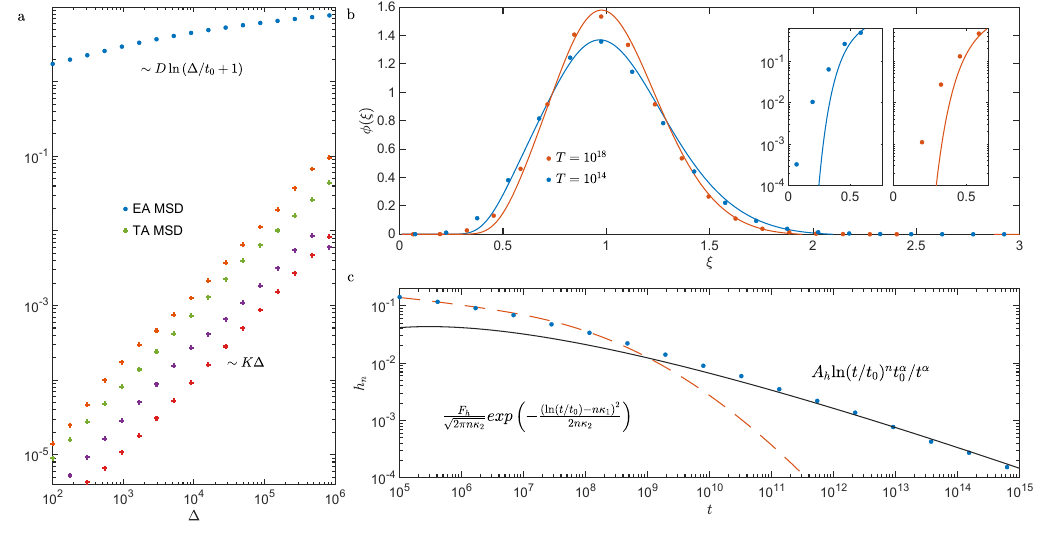}
\caption{\textbf{Ergodicity Breaking in Log-Aging Diffusion.}  
(a) Numerical simulations of the TA and EA MSDs versus lag time \(\Delta\) for \(T=10^8\). The EA-MSD (dots) exhibits logarithmic dynamics, while the TA-MSD ( symbols “\(+\)”) scales as \(\Delta\) with a random factor \(K\)  that varies among different particle trajectories and different color corresponds to TA-MSD of different trajectory.  
(b) Probability distribution \(\phi(\xi)\) of the dimensionless TA-MSD (or the scaling factor \(K/\langle K\rangle\) from panel (a)), for \(\Delta=100\) at \(T=10^{14}\) (blue) and \(T=10^{18}\) (orange). The solid lines denote the skewed log-normal distribution (Eq.~\ref{supplement_h} in the Supplementary Materials). Insets show that, for \(\xi\) between 0 and 0.5, the data deviate from the theoretical line, indicating low probability in this rare region.  
(c) The probability \(h_n\) for \(n=4\) jumps at time \(t\) is shown: at short times, \(h_n\) follows a skewed log-normal distribution form (orange dashed line) (Eq.~\ref{supplement_h} in the Supplementary Materials); at long times, it exhibits a power-law tail with a constant \(A_h\) as observed in simulations(black solid line)(Eq.~\ref{supple_eq_hlong} in the Supplementary Materials). Parameters: \(\alpha=0.5\) and \(t_0=10^2\).}
\label{fig2_eb}
\end{figure*}

To fully understand the ergodicity of ultraslow systems, we focus on the time-averaged (TA) squared displacement\cite{PhysRevLett.101.058101} measured over a time interval \(\Delta\) within a total measurement time \(T\). It is convenient to introduce the time-averaged mean squared displacement (TA-MSD):
\begin{equation}
\overline{\delta^2}(\Delta, T) = \frac{\int_0^{T-\Delta}\left[x\left(t^{\prime}+\Delta\right) - x\left(t^{\prime}\right)\right]^2 dt^{\prime}}{T - \Delta}.
\end{equation}
We simulate trajectories for log-aging diffusion on an unbiased lattice walk. As shown in Fig.~\ref{fig2_eb}a, TA-MSD trajectories ('+' symbols) lie below the ensemble-averaged mean squared displacement (EA-MSD) trajectory (dot symbols) at the same 
observation time \(\Delta\), indicating slower dynamics compared to the ensemble average. Moreover, on a log-log scale the TA-MSD exhibits linear behavior, while the EA-MSD shows increasing curvature, suggesting that its dynamics become slower over time. These marked differences between the TA and EA MSDs reveal ergodicity breaking in log-aging diffusion.

In log-aging diffusion, with memory effect that progressively increase waiting times, the number of jumps \(n\) grows only logarithmically with physical time \(t\) (\(n \propto \ln(t/t_0)\)). Under a matching clock framework\cite{li2024routerandomprocessultraslow}, which synchronizes the internal time (jump number \(n\)), and physical observation time \(t\), the MSD scales as \(x^2\propto n\) (see Supplementary Materials Section A). Ensemble averages capture the overall dynamics of a random process but do not fully characterize its behavior, such as fluctuations among individual trajectories\cite{Krumbeck2021}. For log-aging diffusion, both the mean and variance of the jump number exhibit asymptotic logarithmic time dependence (see Supplementary Materials Section A  and  Fig.~\ref{supplement_fign}),  in contrast to subdiffusion, where the variance scales with the square of the mean.  Consequently, the distribution of log-aging diffusion at finite time converges slowly toward a log-normal distribution (Supplementary Fig.~\ref{supplement_fig.log}).\\

\textbf{The fat-tail singularity beyond the Central Limited Theorem}\\

As shown in Fig.~\ref{fig2_eb}a, the TA-MSDs of four individual trajectories differ, demonstrating that TA-MSDs are random variables. Fig.~\ref{fig2_eb}b plots the probability distribution \(\phi(\xi)\) of the dimensionless TA-MSD, defined as \(\xi = \overline{\delta^2}(\Delta,t)/\langle \overline{\delta^2}(\Delta,t) \rangle\). As time increases, the distribution converges to a skewed log-normal distribution (see detailed derivation in Supplementary Materials Section A), with deviations at small \(\xi\) diminishing. As noted in the ergodicity breaking section, the TA-MSD scales linearly with the jump number \(n\), so that \(\xi = n/\langle n\rangle\). Consequently, \(\xi\) and its distribution \(\phi(\xi)\) can also be interpreted as the probability \(h_n(t)\) of observing \(n\) jumps at time \(t\). However, an exotic asymptotic behavior of probability emerges at large time, far beyond the log-normal distribution.

As shown in Fig.~\ref{fig2_eb}c, although the probability \(h_n(t)\) initially follows a log-normal dependence on \(t\), it later exhibits a slow long-tail decay of the form \(\propto\frac{\left[\ln(t/t_0)\right]^{n}}{\left(t/t_0\right)^{\alpha}}\) (see Supplementary Materials section A and  Fig.~\ref{supplement_fighn2}). This indicates the existence of rare events in log-aging diffusion, such that the probability of events with small jump numbers \(n\) or small \(\xi\) remains higher than predicted by the log-normal distribution. 
Consequently, a fraction of the data lies above the skewed log-normal function (insets in Fig.~\ref{fig2_eb}b). The occurrence of rare events introduces fluctuations in the TA-MSDs.

{\subsection*{Rare events of log-aging process captured by large deviation analysis}}

\begin{figure*}[hbt] 
\centering
\includegraphics[width=1\textwidth]{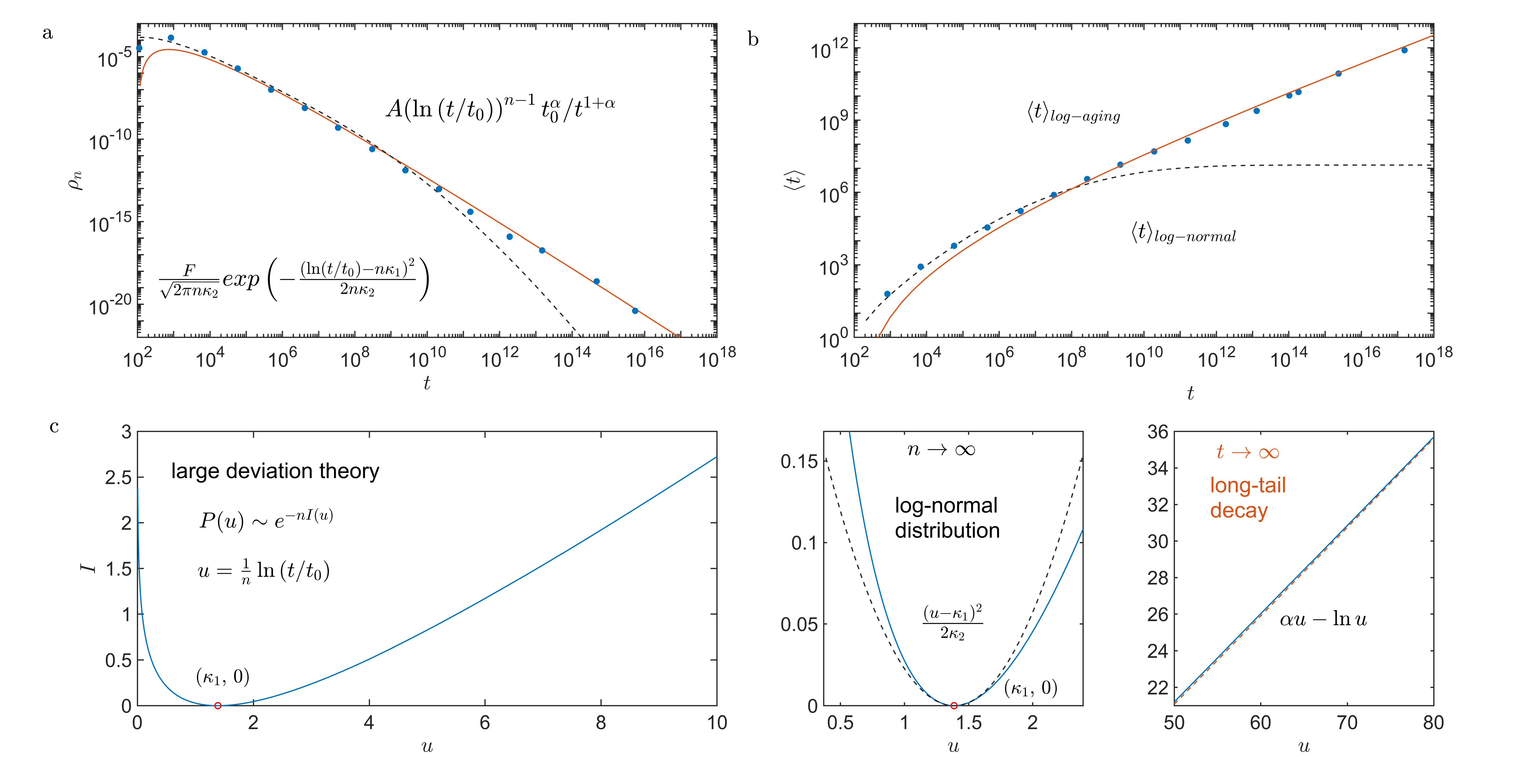}
\caption{   \textbf{Numerical Statistical Properties of Log-Aging Diffusion.} (a) The probability \(\rho_n\) of making \(n\) jumps at time \(t\) follows a log-normal distribution at short times where \(F=\frac{1}{t}\), while at long times it develops a long tail. Parameters: \(\alpha = 0.5\), \(t_0 = 10^2\), \(n = 4\). (b) The influence of rare events on the characteristic time is shown by plotting the time-averaged cutoff \(\langle t\rangle=\int_{t_0}^{t}t^\prime\rho_n(t^\prime)dt^\prime\) versus the cutoff time \(t\). The dashed line represents the time average of the distribution, \(\langle t\rangle_{log-normal}=\int_{t_0}^{t}\frac{1}{\sqrt{2\pi n\kappa_2}}exp\left({-\frac{\left(\ln(t^\prime/t_0)-n\kappa_1\right)^2}{2n\kappa_2}}\right)dt^\prime\) and \(\langle t\rangle_{log-aging}=\int_{t_0}^{t} A{\left(\ln(t^\prime/t_0)\right)^{n-1}t_0^\alpha}/{t^{\prime \alpha}}dt^\prime\), with a constant \(A\). (c) The rate function \(I\) obtained from the large deviation analysis for log-aging diffusion. The logarithm of the time \(t\)  for \(n\) jumps  is expressed as a sum of n variables \(u\), so that the probability of \(\ln{(t/t_0)}/n=u\) scales as \(P(u)\sim e^{-nI} \) (see Supplementary Materials Section C). In the limit \(n\to\infty\), the distribution concentrates around \(\kappa_1\) corresponding to log-normal distribution. Conversely, as \(t\to\infty\) (or \(u\to\infty\)), \(I\)  approaches to a linear form, yielding a slow-decaying heavy tail, highlighting the influence of rare events and results in exotic dynamics in the large time limit.} 
\label{fig3_rho}
\end{figure*}

In log-aging diffusion, these rare events critically dominate the system's dynamics, as illustrated in Fig.~\ref{fig3_rho}a, which shows the evolution of the probability \(\rho_n(t)\) that the walker reaches state \(n\) at time \(t\). Most trajectories concentrate in the easily accessible region (black dashed line), while a few extend into the rare region (orange solid line), exhibiting a fatter tail than the typical log-normal distribution. In this rare region, the probability decays slowly as \(\propto\frac{[\ln(t/t_0)]^{n-1}t_0^\alpha}{t^{1+\alpha}}\) (see Supplementary Materials Section A for details), though the probability here is small, the product of time and probability is enormous and dominates. Consequently, the occurrence of rare events effectively prolongs the system's lifetime, as shown in Fig.~\ref{fig3_rho}b (see also Fig.~\ref{supplement_figrho}). Many physical observations, such as susceptibility, conductance, correlation functions, are influenced by characteristic time of the system\cite{PhysRevB.51.6411}. In log-aging diffusion, the rare events determines the ensemble time and brings unusual dynamics of observations.

The large deviation theory(see Supplementary Materials Section C) provides a powerful mathematical framework for capturing the tail behaviors of probability density functions. In log-aging diffusion, the forward waiting time distribution introduces scaling that varies with the time tick (see Fig.~\ref{fig1}a). As a result, the total time for \(n\) jumps becomes the product of \(n\) independent and identically distributed random variables. This allows us to perform a large deviation analysis on log-aging diffusion and derive the corresponding rate function (details in Supplementary Materials Section C and  Fig.~\ref{supple_plot}).
As shown in Fig.~\ref{fig3_rho}c, the rate function exhibits a local minimum, \(\kappa_1\). Around this point, the rate function is well described by a parabolic approximation. Consequently, as \(n\) increases, the probability distribution concentrates around the minimum and converges to the log-normal distribution. However, far from the minimum point, the rate function displays a linear dependence combined with a logarithmic term. This large deviation analysis reveals a long-tailed decay of the probability for sufficiently long times (see Fig.~\ref{fig2_eb}c and Fig.~\ref{fig3_rho}a), thereby indicating the significant impact of rare events (Table~\ref{table}). Complex structural materials provide an enormous energy landscape that leads to diverse dynamics, such as normal diffusion, subdiffusion, and log-aging diffusion(Table~\ref{table} and Supplementary Materials Section B).

In log-aging diffusion, rare events are highly sensitive to observation because of their exponentially small fraction. For statistical quantities, their impact is negligible, and the log-normal distribution provides an excellent approximation (e.g., Fig.~\ref{fig2_eb}b). However, for dynamical quantities sensitive to rare events—such as the characteristic time—even infrequent occurrences with extra-long durations contribute significantly, and their influence is dominant.\\

{\subsection*{Conclusion}}

\begin{table}[]
\renewcommand{\arraystretch}{1} 

\caption{\textbar\textbf{Comparison of normal, sub-, and log-aging diffusion.} This table summarizes key properties (scaling, distributions, ergodicity). Notably, log-aging diffusion is non-renewal (time-dependent forward waiting time distribution \(\psi_1\)), unlike renewal-based normal and subdiffusion. Log-aging also exhibits distinct asymptotics: in the limits \(n\to\infty\), the probability approximates a log-normal form; while with it \(t\to\infty\) behavior exhibiting a power-law decay modulated by logarithmic correction. The rare events dominate time-averaged quantities such as susceptibility.}
\label{table}\centering 
\begin{tabular}{cccc}
\toprule 
\textrm{Properties}&
\textrm{normal diffusion}&
\textrm{subdiffusion}&
\textrm{log-aging diffusion}\\
\midrule 
\vspace{4mm}
$\psi\quad / \quad\psi_1$ & \( {\tau}^{-1}e^{-t/\tau}\) & \( \sim \frac{A_\alpha}{t^{1+\alpha}}\) & \( \frac{\sin{\pi\alpha}}{\pi}\frac{t_a^\alpha}{(t+t_a)t^\alpha}\)\\
\vspace{4mm}
scaling & breaking & invariance & breaking due to aging \\
\vspace{4mm}
probability distribution & Gaussian & L\'evy &  log-normal  \\
\vspace{4mm}
mean characteristic time & finite & infinite &   divergent following \(\int_0^\infty{A\frac{\left(\ln{(t/t_0)}\right)^{n-1}}{(t/t_0)^\alpha}}dt\) \\
ergodicity & ergodic & breaking & weakly breaking  \\
\bottomrule 
\end{tabular}

\end{table}

We employ log-aging diffusion to model non-equilibrium systems, identifying rare events as a key driver of aging phenomena. Although these events become increasingly infrequent over time, they maintain a significant contribution to the system's dynamics. Specifically, in log-aging diffusion, such rare events result in the characteristic ultraslow dynamics and a novel type of ergodicity breaking. Understanding their crucial role provides valuable insights into non-equilibrium systems, as these events generate heavy-tailed probability distributions (highlighting the relevance to large deviation analysis\cite{PhysRevLett.130.207104}) and may introduce exotic first-passage time dynamics\cite{Barbier-Chebbah2024}, offering methods to quantify the dynamical features and low-frequency noise spectrum of the systems.\\
\printbibliography[segment=\therefsegment, title={References}]
\end{refsegment}

\let\oldaddcontentsline\addcontentsline
\renewcommand{\addcontentsline}[3]{}

\clearpage
\let\addcontentsline\oldaddcontentsline

\newpage

\newpage
{
	\center \bf \large 
	Supplemental Material\\
	
	\vspace*{0.5cm}
}
\begin{refsegment}
\setcounter{equation}{0}
\setcounter{figure}{0}
\setcounter{table}{0}
\setcounter{page}{1}
\renewcommand{\theequation}{S\arabic{equation}}
\renewcommand{\thefigure}{S\arabic{figure}}
\renewcommand{\theHfigure}{Supplement.\thefigure}
\tableofcontents
\clearpage
Unlike conventional anomalous diffusion phenomena, log-aging diffusion exhibits a unique characteristic through its explicit dependence on the initial time. This distinctive process provides a microscopic foundation for understanding logarithmic ultraslow dynamics coupled with aging effects, a phenomenon that has garnered significant attention in experimental studies of complex systems ranging from biological materials\cite{13prl_biology_nature} to physical systems\cite{log_aging_nature,PhysRevLett.118.085501,Ovadyahu_PhysRevLett.92.066801,13prl_glass_doi:10.1073/pnas.1608057113}. Recent studies\cite{li2024routerandomprocessultraslow} have established its connection to glasses and Griffiths-McCoy singularities through scaling analysis of autocorrelation functions, emphasizing its critical role in understanding disordered off-equilibrium systems. This experimental-theoretical synergy underscores the necessity for systematic investigation of log-aging diffusion dynamics.

In this supplement, we first present the derivations of key related quantities—including the time-averaged mean squared displacement (TA MSD), the jump number, and associated probabilities—in Section A. In Section B, we compare log-aging diffusion with normal diffusion and subdiffusion, highlighting that, although rare events become increasingly infrequent in log-aging diffusion, their presence significantly influences the overall process.  In Section C, we give a large derivation viewpoint to analyze the probability distribution and capture the rare events. 
Fig. ~\ref{supplement_fign} gives statistic information.
Fig. ~\ref{supplement_fighn2} and  \ref{supplement_figrho} illustrate the presence of rare events in log-aging diffusion and their impact on the characteristic time of the process. Meanwhile, Fig. ~\ref{supplement_fig.sub} and \ref{supplement_fig.log} demonstrate the scale-invariance properties of subdiffusion and log-aging diffusion, revealing two distinct manifestations of rare events and the corresponding dynamical behaviors. Fig.~\ref{supple_plot} shows large derivation theory for log-aging diffusion.

\section{ Derivations of Related Quantities in the Main Text}
For log-aging diffusion, the random process\cite{PhysRevLett.110.208301} based on the continuous-time random walk (CTRW) framework provides a vivid description of the process.
The ergodicity breaking in log-aging diffusion necessitates trajectory-dependent analysis through time-averaged mean squared displacement (TA MSD). Within continuous-time random walk formalism, the step number $n$ serves as a trajectory-specific internal clock, where both variance $\langle \Delta n^2 \rangle$ and mean $\langle n \rangle$ scale identically with $\ln(t/t_0)$, suggesting log-normal statistics. However, the step-counting probabilities $h_n(t)$ (probability for $n^\text{th}$ step at $t$) and $\rho_n(t)$ (probability of completing $n$ steps by $t$) exhibit dynamical crossovers: initial log-normal profiles transition to long-tail decays. This $n$-dependent crossover, evidenced by infinite-lifetime rare events, fundamentally links aging effects to extreme-value statistics in disordered systems.


\subsection{ Time-Averaged MSD}

For the clock matching process in diffusion\cite{clock_PhysRevE.71.026101,li2024routerandomprocessultraslow}, the mean squared displacement scales with the jump number, i.e., in internal time (with jump number \(n\)), diffusion is a simple Markov process so that \(x^2\sim n\). When observed in physical time, one obtains \(x^2(t)\sim n(t)\)\cite{PhysRevLett.101.058101,doi:10.1142/9789814340595_0013}. This matching underlies diffusion irrespective of whether ensemble or time averages are considered. Thus, the time-averaged MSD satisfies \(\bar{\delta}^2(\Delta,t)\propto n(t)\).

Assuming\cite{PhysRevLett.101.058101}
\[
\left[x\left(t'+\Delta\right)-x\left(t'\right)\right]^2\sim\langle\delta x^2(t)\rangle\, n\left(t'+\Delta,t'\right),
\]
where 
\[
n\left(t'+\Delta,t'\right)=n\left(t'+\Delta,0\right)-n\left(t',0\right)
\]
denotes the number of jumps between times \(t'\) and \(t'+\Delta\), then in the limit \(T\gg\Delta\) we obtain
\begin{equation}\label{supplement_TAMSD}
\bar{\delta}^2(\Delta,t) \sim \langle\delta x^2(t)\rangle\, \frac{\Delta}{t}\, n(t).
\end{equation}

\subsection{Jump Number}

We denote the probability that the walker makes \(n\) jumps by time \(t\) as \(h_n(t)\). Its Mellin transform is given by\cite{PhysRevLett.110.208301}

\begin{equation}
\label{supple.hnp}
h_n(p)\equiv \int_0^\infty t^{p-1} \rho(t)\, dt =t_0^p\frac{G(p)-1}{p}\, G^n(p).
\end{equation}
Thus, the average jump number in Mellin space is

\begin{equation}
\label{supplement_n}
\langle n(p)\rangle = \sum n\, h_n(p) = t_0^p\frac{G(p)-1}{p}\, G(p)\frac{d}{dG(p)} \sum G^n(p) = t_0^p\frac{G(p)}{p\left[G(p)-1\right]}.
\end{equation}
Similarly, the second moment is

\begin{equation}
\label{supplement_n2}
\langle n^2(p)\rangle = \sum n^2\, h_n(p) =t_0^p \frac{G(p)-1}{p}\, \Biggl\{ G(p)\frac{d}{dG(p)} \Biggr\}^2\sum G^n(p) =-t_0^p \frac{G(p)\left[1+G(p)\right]}{p\left[G(p)-1\right]^2}.
\end{equation}
Given 
\[
G(p)=\frac{\sin(\pi\alpha)}{\pi}\frac{\Gamma(\alpha-p)}{\Gamma(1-p)}\, (0<\alpha<1),
\]
the asymptotic expansions of Eqs.~\ref{supplement_n} and \ref{supplement_n2} in the long-time limit are
\begin{equation}
\begin{aligned}
\langle n(t)\rangle &\sim \frac{\ln(t/t_0)}{\kappa_1},\\[1mm]
\langle  n^2(t)\rangle & \sim \frac{[\ln(t/t_0)]^2}{\kappa_1^2} + \frac{\kappa_2}{\kappa_1^3}\ln(t/t_0),
\end{aligned}
\end{equation}
where \(\kappa_n=\frac{d^n}{d\alpha^n}\ln G(p)\big|_{p=0}\). It is evident that both the average and variance, \(\langle \delta^2 n(t)\rangle = \langle (n(t)-\langle n(t)\rangle)^2\rangle\), of the jump number scale linearly with \(\ln(t/t_0)\), Fig.~\ref{supplement_fign}a.

\subsection{ Statistical Characteristics of TA MSD}

Eq.~\ref{supplement_TAMSD} linearly relates the time-averaged mean squared displacement (TA MSD), \(\overline{\delta^2}(\Delta,t)\), to the jump number \(n(t)\). In order to investigate TA MSD, it is necessary to analyze the statistical distribution of the jump number.

The Mellin transform of \(h_n(p)\) (Eq.~\ref{supple.hnp}) exhibits singularities at \(p=\alpha+j-1\) for \(j=1,2,3,\dots\). Using the properties of the \(\Gamma\) function\cite{lens.org/122-254-545-854-246}, the inverse Mellin transform can be performed along the right infinite semicircle:
\[h_n(t)  = \frac{1}{2\pi i} \int_{c-i\infty}^{c+i\infty} t^{-p} h_n(p)\,d p 
    = - \sum_{j=1} \operatorname{Res}\{h_n(p)t^{-p},p=\alpha+j-1\},\]
where
\[\operatorname{Res}\{h_n(p)t^{-p},p=\alpha+j-1\}= \frac{1}{(j-1)!}{\frac{d^{j-1}}{dp^{j-1}} \left[(p-\alpha-j+1)^n  \frac{h_n(p)}{t^{p}}\right]}\Bigg|_{p=\alpha+j-1}.\]  
In the limit \(t\to\infty\), the residue at \(p=\alpha\) dominates, and the asymptotic behavior of \(h_n(t)\) is given by
\begin{equation}\label{supple_eq_hlong}
   h_n(t)\sim A_h\frac{\left[\ln(t/t_0)\right]^{n}t_0^\alpha}{t^\alpha}, \quad t>t^\star,
\end{equation}
where \(A_h\) is related to \(n\) and \(t^\star\) depends on \(n\); larger \(n\) implies a larger \(t^\star\). This long-tail decay leads to a divergent mean waiting time, indicating the occurrence of rare events.

In the limit \(n\to\infty\), \(h_n(t)\) takes the form\cite{PhysRevLett.110.208301}:
\begin{equation}\label{supplement_h}
h_{n}(t)=\frac{\kappa_1}{\sqrt{2 \pi \kappa_2 n}} \exp\left(-\frac{\left[\ln\left(t/t_0\right)-\kappa_1 n\right]^2}{2 \kappa_2 n}\right) \, F\!\left(\frac{\ln\left(t/t_0\right)-\kappa_1 n}{\sqrt{\kappa_2 n}}\right),
\end{equation}
where the polynomial 
\[
F(x)=1+\frac{\kappa_2+\kappa_2^2}{2\kappa_2\sqrt{\kappa_2 n}}\, x+\frac{\kappa_3 n}{6\left(\kappa_2 n\right)^{3/2}}\left(x^3-3x\right)+\dots
\]
accounts for higher-order corrections. This expression represents a skewed log-normal distribution, reflecting that the time \(t\) for \(n\) jumps is effectively the product of random variables\cite{PhysRevLett.110.208301}.

Simulations of the probability \(h_n(t)\) for the \(n^\text{th}\) jump (Fig.~\ref{supplement_fighn2}) show that the log-normal distribution provides a good approximation at early times, while a pronounced long-tail decay emerges at later times. Moreover, as \(n\) increases, the transition from log-normal behavior to long-tail decay occurs over a longer time scale. This suggests that the relative contribution of rare events diminishes over time, leading to fewer jump events as time progresses.

By combining Eq.~\ref{supplement_TAMSD} and Eq.~\ref{supplement_h}, the probability distribution for the dimensionless TA MSD,
\[
\xi = \frac{\overline{\delta^2}(\Delta,t)}{\langle \overline{\delta^2}(\Delta,t) \rangle},
\]
is obtained as
\begin{equation}
\label{supple_xi}
\phi(\xi) \sim \sqrt{\frac{\kappa_1\ln(t/t_0)}{2\pi \kappa_2 \xi}}\, \exp\left(-\frac{\kappa_1 \ln(t/t_0)(\xi-1)^2}{2\kappa_2}\right) \, F\!\left(\xi,\ln(t/t_0)\right),
\end{equation}
with
\[
F\left(\xi,\ln(t/t_0)\right)=1+\left[\frac{\kappa_2+\kappa_1^2}{2\kappa_2\xi} - \frac{\kappa_3 \kappa_1}{2\kappa_2^2\xi}\right](1-\xi)+\frac{\kappa_3 \kappa_1^2}{6\kappa_2^3\xi^2}(1-\xi)^3\ln(t/t_0)+\dots.
\]
As noted above, the presence of rare events increases the probability for small \(n\), leading to deviations in Eq.~\ref{supple_xi} at small \(\xi\), Fig.~\ref{supplement_fign}c.

Finally, the variance of \(\xi\) for a lag time \(\Delta\) measured over an experimental time \(T\) (\(\Delta\ll T\)) is given by (Fig.~\ref{supplement_fign}b)
\begin{equation}
    \frac{\langle \delta \xi^2 \rangle}{\langle \delta \xi \rangle^2} = \frac{\langle \left[\overline{\delta^2}(\Delta,T) - \langle \overline{\delta^2}(\Delta,T)\rangle\right]^2 \rangle}{\langle \overline{\delta^2}(\Delta,T)\rangle^2} = \frac{\langle (n - \langle n\rangle)^2 \rangle}{\langle n\rangle^2}\sim \frac{\kappa_2}{\kappa_1 \ln (T/t_0)}.
\end{equation}
Interestingly, this coincides with the variance of the jump number reported in \cite{PhysRevLett.110.208301}, resulting from the matching between internal and physical time scales in Eq.~\ref{supplement_TAMSD}.

The ergodicity-breaking (EB) parameter\cite{PhysRevLett.101.058101} (also called the heterogeneity parameter \cite{PhysRevE.89.012136}), defined as 
\[
\text{EB} = \lim_{T\to\infty} \frac{\langle \delta \xi^2 \rangle}{\langle \delta \xi \rangle^2},
\]
which quantifies the fluctuations of the dimensionless TA-MSD. This parameter decreases as the observation time \(t\) increases for log-aging diffusion (Fig. \ref{supplement_fign}), indicating a gradual reduction in the fraction of rare events over time. In log-aging diffusion, a complex energy landscape with rare events reinforces memory effects. Over time, the overall dynamics progressively slow down, and the occurrence of extremely long waiting times—longer than those of most trajectories—becomes rare. Thus, the skewed log-normal distribution fits better for longer observation times, that is larger jump number \(n\)(Fig.~\ref{supplement_fig.log}), indicating a reduced proportion of small \(\xi\) events. Although rare events are infrequent, they can have significant effects. 
It is now clear that the magnitude of trajectory fluctuations (EB) alone is insufficient to define ergodicity breaking; this suggests that rare event statistics by themselves are inadequate. We define the parameter 
\begin{equation}
\label{eq_eta}
\eta= \lim\limits_{t\to\infty}\frac{\left[\overline{\delta^2}(\Delta, t)-\langle x^2(\Delta)\rangle\right]^2}{\langle x^2\rangle^2(\Delta)},
\end{equation}
displaying discrepancy between them( log-aging diffusion is shown in Fig.~\ref{supplement_fign}b). For ergodic process, like normal diffusion, time average of long trajectory converges to ensemble average, thus $\eta=0$. For non-ergodic process, $\eta\ne0$.

\subsection{ Rare Events}

Although rare events occur infrequently, their impact is significant. For example, the sum of random variables is often dominated by a few exceptionally high values. Below, we illustrate this effect in log-aging diffusion. The probability \(\rho_n(t)\) that a passenger arrives at the \(n^\text{th}\) stop at time \(t\) is given by

\begin{equation}
\rho_n(t)=\int_0^t \rho_{n-1}(t')\, \psi_1\left(t-t' \mid t'\right) dt',
\end{equation}
where \(\psi_1\left(t-t'\mid t'\right)\) denotes the forward waiting time distribution. As shown in Fig.~\ref{supplement_figrho}, \(\rho_n(t)\) exhibits a power-law decay for \(t>t^\star\), with the time \(t^\star\) increasing as \(n\) becomes larger. Since the total time \(t_n\) for \(n\) jumps is the sum of \(n\) random variables, the long-tail decay reveals the presence of rare events that, although unlikely, have a profound influence on the system. Moreover, as the cutoff time  in the average

\[
\langle t \rangle=\int_0^{t} t^\prime\,\rho_n(t^\prime)\,dt^\prime
\]
increases, the computed mean time grows and converges to the average obtained from the long-tail decay, \(\langle t \rangle_{log-aging}\)(See Fig.~\ref{supplement_figrho}). This clearly demonstrates that rare events dominate the system's dynamics.

\subsection{ Relation Between the Asymptotic Expressions of \texorpdfstring{\(\rho_n\)} and \(h_n\)}

The probability \(h_n(t)\) that a walker makes its \(n^\text{th}\) jump at time \(t\) can be expressed as the product of two factors: (i) the probability \(\rho_n(t')\) that the walker has completed \(n\) jumps by time \(t' < t\) and (ii) the probability that no jump occurs between \(t'\) and \(t\). Hence, we write

\begin{equation}
h_n(t)=\int_0^t \rho_n(t')dt'-\int_0^t \rho_{n+1}(t')dt'=\int_0^t \rho_n(t') \left[\int_{t-t'}^\infty \psi_1(\tau\mid t')\,d\tau\right] dt'.
\end{equation}
 Their Mellin transforms are related by\cite{PhysRevLett.110.208301}

\[
h_n(p)=\frac{G(p)-1}{p}\,\rho_n(p+1).
\]
Expanding for small \(p\), \(h_n(p)\sim t_0^p\kappa_1e^{n\kappa_1p}\, \text{and}\, \rho_n(p+1)\sim t_0^pe^{n\kappa_1p}\), gives 
\[h_n(p)/t_0^p\sim \frac{d}{ndp}\left(\rho_n(p+1)/t_0^p\right),\]
which implies—via the properties of the Mellin transform—that for large \(t\) one obtains

\[
nh_n(t)\sim t\ln{(t/t_0)}\rho_n(t).
\]
Although \(h_n(t)\) and \(\rho_n(t)\) share similar asymptotic behavior, they are normalized differently, which can be explained by the using of their Mellin transforms.

For the normalization of $h_n(t)$\cite{PhysRevLett.110.208301}:
summing over all $n$ terms of Eq.~\ref{supple.hnp} yields
\begin{equation*}
    \sum\limits_{n=0}h_n(p) = t_0^p \frac{G(p)-1}{p}\sum\limits_{n=0} G(p)^n = -\frac{t_0^p}{p}
\end{equation*}
Since the Mellin inversion of $-{t_0^p}/{p}$ is $H(t-t_0)$ (\(H\) is Heaviside function), we obtain
\begin{equation*}
    \sum_{n=0}h_n(t) = 1
\end{equation*}
For the normalization of $\rho_n(t)$\cite{PhysRevLett.110.208301}:
\[
 \rho_n(p) \equiv \int_0^\infty t^{p-1} \rho(t)\, dt = t_0^{p-1}G(p-1)^n
\]
 Noting that $G(0)=1$, we can obtain 
\begin{equation*}
    \int_0^\infty \rho_n(t)\, dt =\rho_n(p=1)= 1
\end{equation*}

When considering observables related to the jump number—such as the particle's positional probability—\(h_n\) is the appropriate measure. Conversely, for time-related observables—such as the first passage time in a target problem—\(\rho_n\) should be used. Therefore, it is necessary to investigate these probabilities independently.

\section{ Comparison with Renewal Diffusion}

In conventional continuous-time random walk (CTRW) models—applicable to both normal diffusion and subdiffusion—the process is renewal, meaning that the waiting time distribution is identical and independent for each jump.
\subsection{ Normal Diffusion}
For normal diffusion, the waiting time distribution has a finite first moment. For example, one may use
\[
\psi(t)=\frac{1}{\tau}\exp\left(-\frac{t}{\tau}\right),
\]
so that the mean waiting time is \(\tau\). In this case, the probability \(\rho_n(t)\) that the walker arrives at the \(n^\text{th}\) stop at time \(t\) is given by
\begin{equation}
\rho_n(t)=\int_0^t \rho_{n-1}(t_{n-1})\,\psi(t-t_{n-1})\,dt_{n-1}
=\frac{t^{n-1}}{(n-1)!\,\tau^n}e^{-t/\tau},
\end{equation}
with \(\langle t\rangle=n\tau\). Similarly, the probability \(h_n(t)\) that the walker makes its \(n^\text{th}\) jump at time \(t\) is
\begin{equation}
h_n(t)=\int_0^t \rho_n(t')\left[\int_{t-t'}^\infty \psi(\tau)\,d\tau\right]dt'
=\frac{t^{n}}{n!\,\tau^n}e^{-t/\tau}.
\end{equation}
Noting that
\[
t\,\rho_n(t)= n\,h_n(t),
\]
we see that the rapid decay of these distributions—with a finite mean—precludes the occurrence of rare events.

\subsection{Subdiffusion}
In contrast, for subdiffusion the waiting time distribution exhibits a long tail,
\[
\psi(t)\sim\frac{A_\alpha}{t^{1+\alpha}},
\]
which is scale invariant and self-similar (see Fig.~\ref{fig1} in the main text), thereby governing the overall dynamics. The asymptotic Laplace transform, at small \(s\), of the probability \(\rho_n(t)\) for arriving at the \(n^\text{th}\) stop behaves as\cite{10.1093/acprof:oso/9780199234868.001.0001}
\[
\rho_n(s)\sim \exp\Bigl(-An\,s^\alpha\Bigr),
\]
so that its time-domain expression for long time can be written as
\begin{equation}
\rho_n(t)=\frac{1}{ A^{1/\alpha} n^{1/\alpha}}\, l_\alpha\!\left(\frac{t}{A^{1/\alpha} n^{1/\alpha}}\right),
\end{equation}
where \(l_\alpha\) is the one-sided Lévy function \cite{Amir_2020} and \(l_\alpha(x)\sim\frac{1}{x^{1+\alpha}}\) for \(x\gg1\). Similarly, the asymptotic Laplace transform of \(h_n(t)\) behaves as
\[
h_n(s)\sim A\,s^{\alpha-1}\exp\Bigl(-An\,s^\alpha\Bigr),
\]
yielding\cite{PhysRevLett.101.058101}
\begin{equation}
h_n(t)=\frac{t}{\alpha A^{1/\alpha} n^{1+1/\alpha}}\, l_\alpha\!\left(\frac{t}{A^{1/\alpha} n^{1/\alpha}}\right).
\end{equation}
In the small-\(s\) limit one can show that \(h_n(s)\sim\frac{d\rho_n(s)}{\alpha nds}\), which implies—by properties of the Laplace transform—that
\[
nh_n(t)\sim t\,\rho_n(t)/\alpha.
\]
This comparison highlights the fundamental difference between normal diffusion—with its rapid, exponential decay and finite moments—and subdiffusion, where the heavy-tailed waiting time distribution leads to scale invariance and long-time dynamics dominated by rare events.

As shown in Fig.~\ref{supplement_fig.sub}, the probability distribution \(h_n(t)\) in subdiffusion exhibits scale invariance; that is, when the jump number \(n\) is appropriately rescaled, the distributions at different times collapse onto a single curve. This behavior implies that a constant fraction of rare events persists throughout the process. Notably, the variance \(\langle \delta^2 n \rangle\) and the mean \(\langle n \rangle\) of the jump number scale with time as
\[
\langle \delta^2 n \rangle\propto t^{2\alpha} \quad \text{and} \quad \langle n \rangle\propto t^{\alpha},
\]
so that the relative fluctuation
\[
\frac{\langle \delta^2 n \rangle}{\langle n \rangle^2}
\]
approaches a nonzero constant. Consequently, the ergodicity breaking parameter serves as an effective measure for systems with scale invariance.

In contrast, for log-aging diffusion the scale invariance is lost and the probability distribution no longer exhibits self-similarity. In this case, both the variance and the mean scale with time as
\[
\langle \delta^2 n \rangle\propto \ln\left({t}/{t_0}\right) \quad \text{and} \quad \langle n \rangle\propto \ln\left({t}/{t_0}\right).
\]
As a result, when the distribution is rescaled by \(n/\langle n \rangle\), its width narrows (see Fig.~\ref{supplement_fig.log}), and the ergodicity breaking parameter approaches zero. Thus, this parameter fails to capture ergodicity breaking in log-aging diffusion, where the diminishing fraction of rare events no longer significantly influences the rescaled dynamics.

These observations underscore the limitations of the conventional ergodicity breaking parameter in systems without scale invariance. Although rare events become progressively less probable in log-aging diffusion, their existence remains critical; even infrequent rare events can dominate the dynamics, offering important insights into systems governed by such extreme fluctuations.

\section{ Large Derivation Theory }
The large deviation theory (LDT) is an important mathematical tool, and it adequately describes the behaviors of the PDF in its tails. In  Cramérs theorem, a sample mean of independent and identically distributed (IID) random variables has the simple form of the scaled cumulant generating function as\cite{TOUCHETTE20091}
\begin{equation}
    \lambda(k)=\ln \left\langle\mathrm{e}^{k X}\right\rangle.
\end{equation}
If \(\lambda\) is exists and is differentiable, then the mean of IID random variables, \(S_n=\frac{1}{n}\sum X_i\), satisfies a large deviation, that the probability is the exponential approximation, as  
\begin{equation}
    P(S_n=x)\sim e^{-nI(x)}
\end{equation}
where \(I(x)\) is rate function, giving a direct and detailed picture of the deviations or fluctuations of \(S_n\) around its typical value. And \(\lambda(k)\) and \(I(x)\) satisfies Legendre Fenchel transforms:
\begin{equation}
\begin{aligned}
  I(x)&=\sup _{k}\{k x-\lambda(k)\} \\
   \lambda(k)&=\sup _{x}\{k x-I(x)\}.
\end{aligned}\label{supple_Ilam}
\end{equation}

In the framework of continuous-time random walks (CTRW), the waiting times in log-aging diffusion are assumed to follow the forward waiting time distribution
\[\psi_1(\tau\mid\,t_a)=\frac{\sin{(\pi\alpha)}}{\pi}\frac{t_a^\alpha}{\tau^\alpha(\tau+t_a)}\]
By introducing the new variable \(X=\frac{\tau+t_a}{t_a}\), the variable \(X\) become independent and identically distributed  with probability density function 
\[\psi_1(x)=\frac{\sin{(\pi\alpha)}}{\pi}\frac{1}{x(x-1)^\alpha}\]
And the time for jump \(n\) steps is the product of \(X\), 
\[t_n=t_0\prod \limits_{i}X_i.\]
Taking the logarithm leads to
\[\ln{t/t_0}=\sum\limits_{i}\ln{X_i}\]
so that the logarithm of the total time is expressed as a sum of i.i.d. random variables. Defining 
\[u=\ln{X},\]
its distribution is
\begin{equation}\label{supple_psiu}
   \phi(u)=\frac{\sin{(\pi\alpha)}}{\pi}\frac{1}{(e^{u}-1)^\alpha}. 
\end{equation}
Next, we compute the scaled cumulant generating function
\begin{equation}
    \lambda(k)=\ln \left\langle\mathrm{e}^{k u}\right\rangle=\ln{\frac{\Gamma(\alpha-k)}{\Gamma(\alpha)\Gamma(1-k)}}\,\,\,\, k<\lambda.
\end{equation}
which is manifestly convex. According to the Legendre--Fenchel transform, for a convex function $\lambda(k)$ the rate function $I(u)$ satisfies \cite{TOUCHETTE20091}
\begin{equation}\label{supple_inversefunction}
    \frac{dI}{du(k)}=k(u)\,\,\,\,\,\,\,\,\,\,\frac{d\lambda}{dk(u)}=u(k).
\end{equation}
 Using the Legendre--Fenchel transform, we can obtain the rate function  \(I(u)\)  from \(\lambda(k)\)
through the following steps:
\begin{enumerate}
\item[(1)] First, we obtain the function \(\lambda(k)\), Fig.~\ref{supple_plot}a.
\item[(2)] Next, we compute the derivative \(\lambda^\prime(k)\). And we find that $\lambda'(k)>0$, the blue curve in Fig.~\ref{supple_plot}b.
\item[(3)] Since the derivative of $\lambda$ and the derivative of $I$ are functional inverses, Eq.~\ref{supple_inversefunction}, the graph of $I'(u)$ can be obtained from that of $\lambda'(k)$ by a reflection about the line $y=x$, the orange dashed curve in Fig.~\ref{supple_plot}b. In doing so, it is found that the variable $u>0$.
\item[(4)] Finally, using the derived $I'(u)$, one can reconstruct the approximate shape of $I(u)$, Fig.~\ref{fig3_rho}c in the main text; notably, one observes that $I(\kappa_1)=0$ and $I'(\kappa_1)=0$.
\end{enumerate}
As illustrated in Fig.~\ref{fig3_rho}c in the main text, the rate function $I(u)$ exhibits a parabolic form only in the vicinity of $\kappa_1$ and is asymmetric about $u=\kappa_1$. Consequently, the small fluctuations of $\ln(t_n/t_0)$ near its typical value are governed by a skewed Gaussian distribution (see Eq.~\ref{supplement_h}).
An alternative representation for $\lambda(k)$ is given by
\[\lambda=\ln{\left(\frac{\sin{(\pi\alpha)}}{\pi}\sum\limits_{i=0}\frac{\Gamma(\alpha+i)}{\Gamma(\alpha)\Gamma(1+i)}\frac{1}{\alpha+i-\lambda}\right)}.\]
Here, the singularity of $\lambda(k)$ at $k=\lambda$ leads to the asymptotic rate function for large \(u\to\infty\) as
\[I(u)\sim\alpha u-\ln{u}.\]
As a consequence, large deviations of the random variable $\frac{1}{n}\sum\limits_i u_i$ away from its typical value decay exponentially. However, when translated back to the original variable $t_n$, the tail behavior becomes even slower, following a power-law with a logarithmic correction, Eq. ~\ref{supple_eq_hlong}. Interestingly, when we examine quantities that mainly reflect statistical averages, the log-normal distribution provides an excellent approximation. For example, the jump number, \(\langle n\rangle= \ln{(t/t_0)}/\kappa_1\), is a natural consequence of the log-normal distribution. However, when considering quantities that are sensitive to rare events—such as the characteristic time—even though rare events occur infrequently, their extra-long durations make their overall contribution significant. In this case, the influence of rare events cannot be ignored. As discussed above, we identify a new kind of rare event: while the frequency of these events diminishes over time, their impact on the process remains substantial.

For normal diffusion with an exponential waiting time distribution, as studied in \cite{PhysRevE.103.042116}, the central limit theorem guarantees that large deviations decay exponentially, so the fast-decaying tail has little influence on the observed dynamics. In contrast, in log-aging diffusion the slow, power-law decay arising from large deviations governs the system dynamics (see Fig.~\ref{supplement_figrho}). Although the tail represents only a small fraction of events, its persistent presence can significantly affect the lifetime of the system, thereby hinting at the occurrence of rare events.

In the subdiffusive case, scale invariance implies that the corresponding distribution asymptotically obeys a scaling law and that a large deviation perspective has been developed \cite{PhysRevLett.130.207104}. Meanwhile, an asymptotic power-law form for the rate function is derived in \cite{PhysRevE.103.042116}.
\printbibliography[segment=\therefsegment, title={References}, filter=notinsegment1]
\end{refsegment}
\renewcommand{\thefigure}{S\arabic{figure}}
\clearpage
\begin{figure}[t]
\includegraphics[width=1\linewidth]{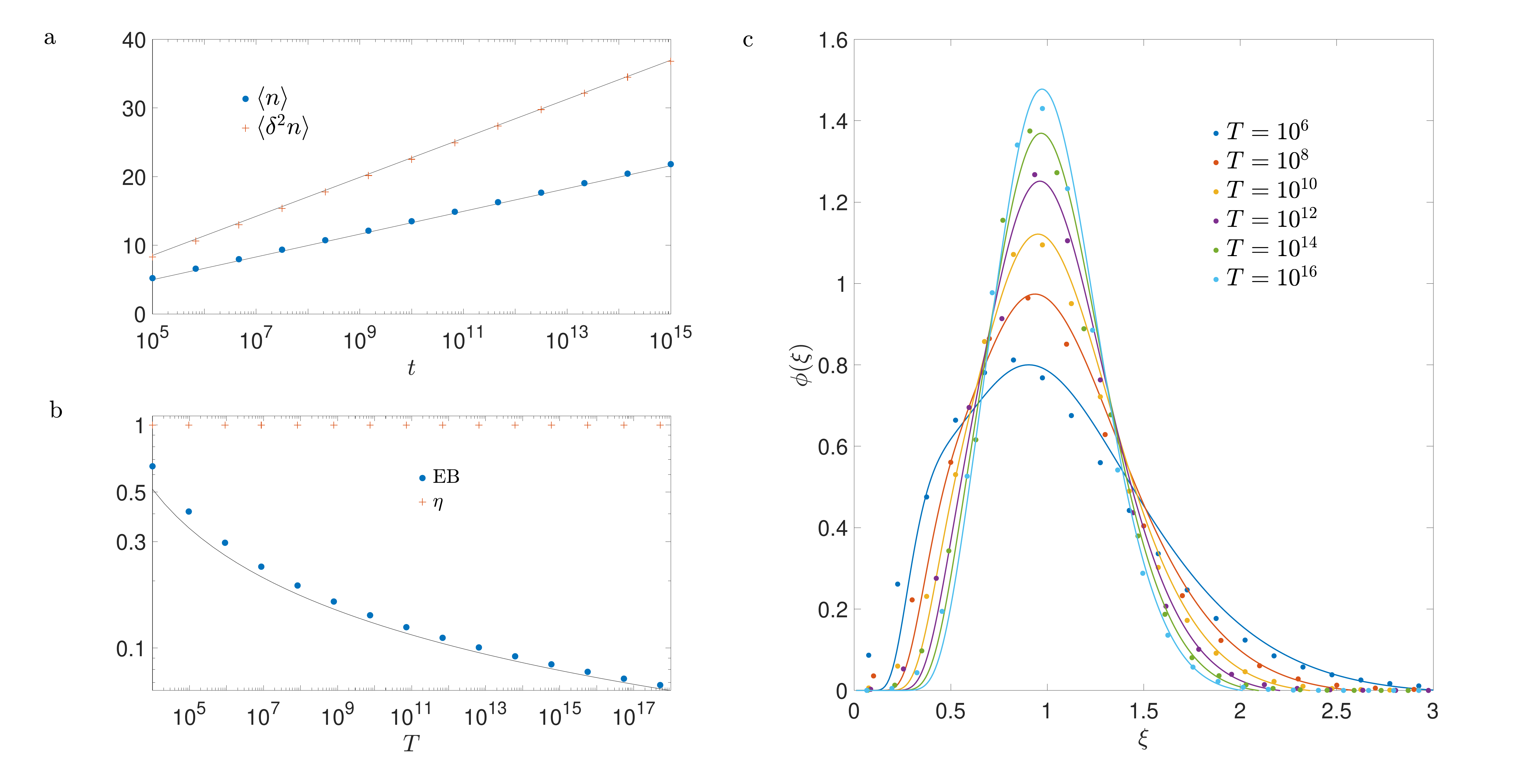}
\caption{(a)Numerical calculations of the statistical jump number at observation time \(t\): \(\langle n\rangle\) is the ensemble average and \(\delta^2n\) its variance. The dashed line indicates the asymptotic behavior, with both axes displayed logarithmically. (b)Parameters EB and \(\eta\) characterize ergodicity at \(\Delta=100\). The EB data (blue dots) decay slowly with logarithmic time, following an asymptotic trend of \(\propto \frac{1}{\ln(t/t_0)}\) (black dashed line), while \(\eta\) (orange '\(+\)') approaches 1, highlighting the discrepancy between time-averaged and ensemble-averaged MSD. (c) The probability of dimensionless TA MSD at different time, \(\Delta=100\).Parameters: \(\alpha=0.5\) and \(t_0=10^2\). }
\label{supplement_fign}
\end{figure}

\begin{figure}[b]
\includegraphics[width=1\linewidth]{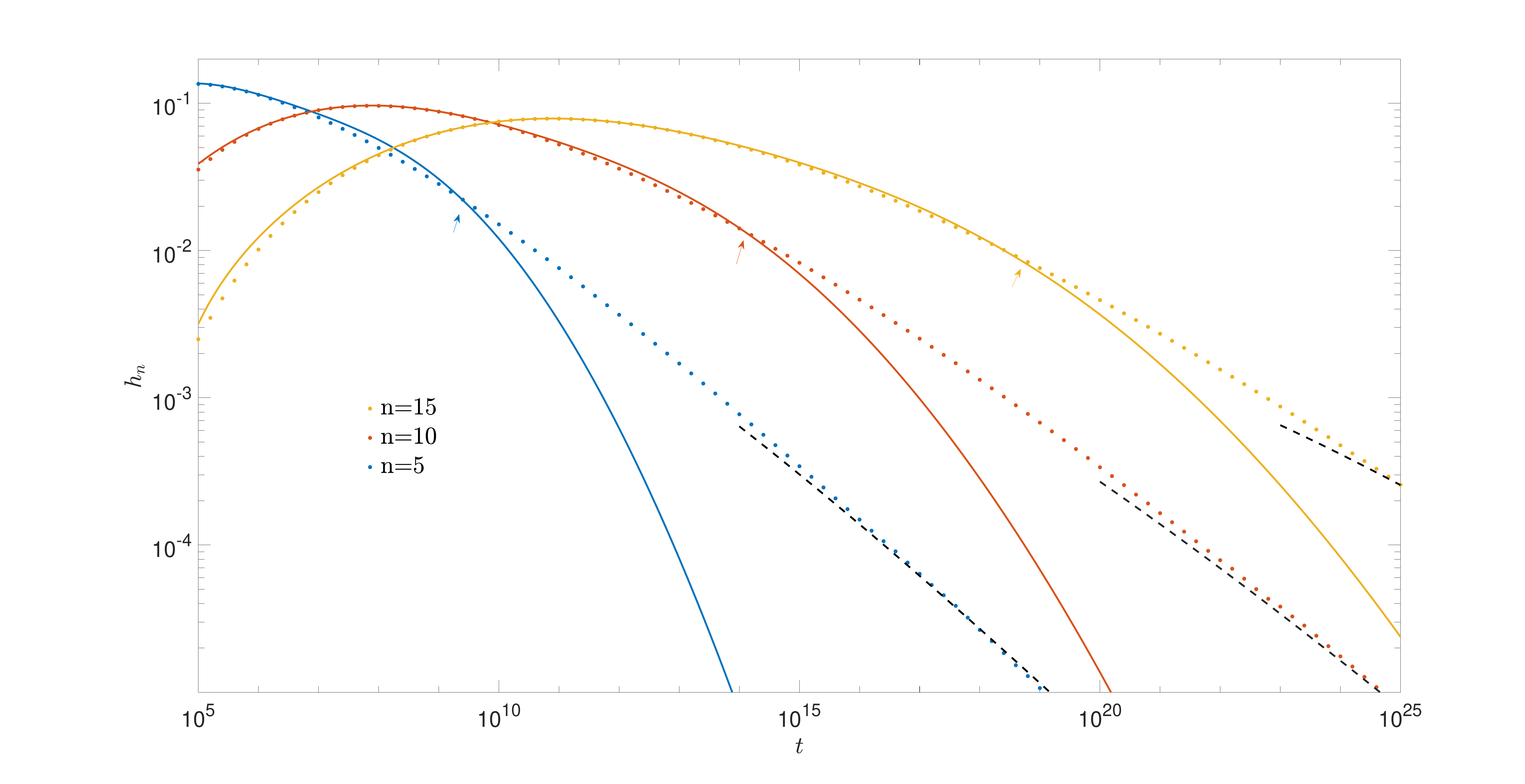}
\caption{\textbf{The existence of rare events in log-aging diffusion.} The probability \(h_n\) for different \(n\)  vs time \(t\) is plotted in log-log axis, with long-tail decay (the dashed black line). The solid lines are depicted with log-normal distribution \(\frac{1}{t\sqrt{2\pi n\kappa_2}}e^{-\frac{\left(\ln(t/t_0)-n\kappa_1\right)^2}{2n\kappa_2}}\), and the dashed lines are with long tail \(A\frac{\left(\ln(t/t_0)\right)^{n-1}t_0^\alpha}{t^{1+\alpha}}\), with constant \(A\). Parameters: \(\alpha=0.5\), \(t_0=100\).}
\label{supplement_fighn2}
\end{figure}

\begin{figure}[b]
\includegraphics[width=1\linewidth]{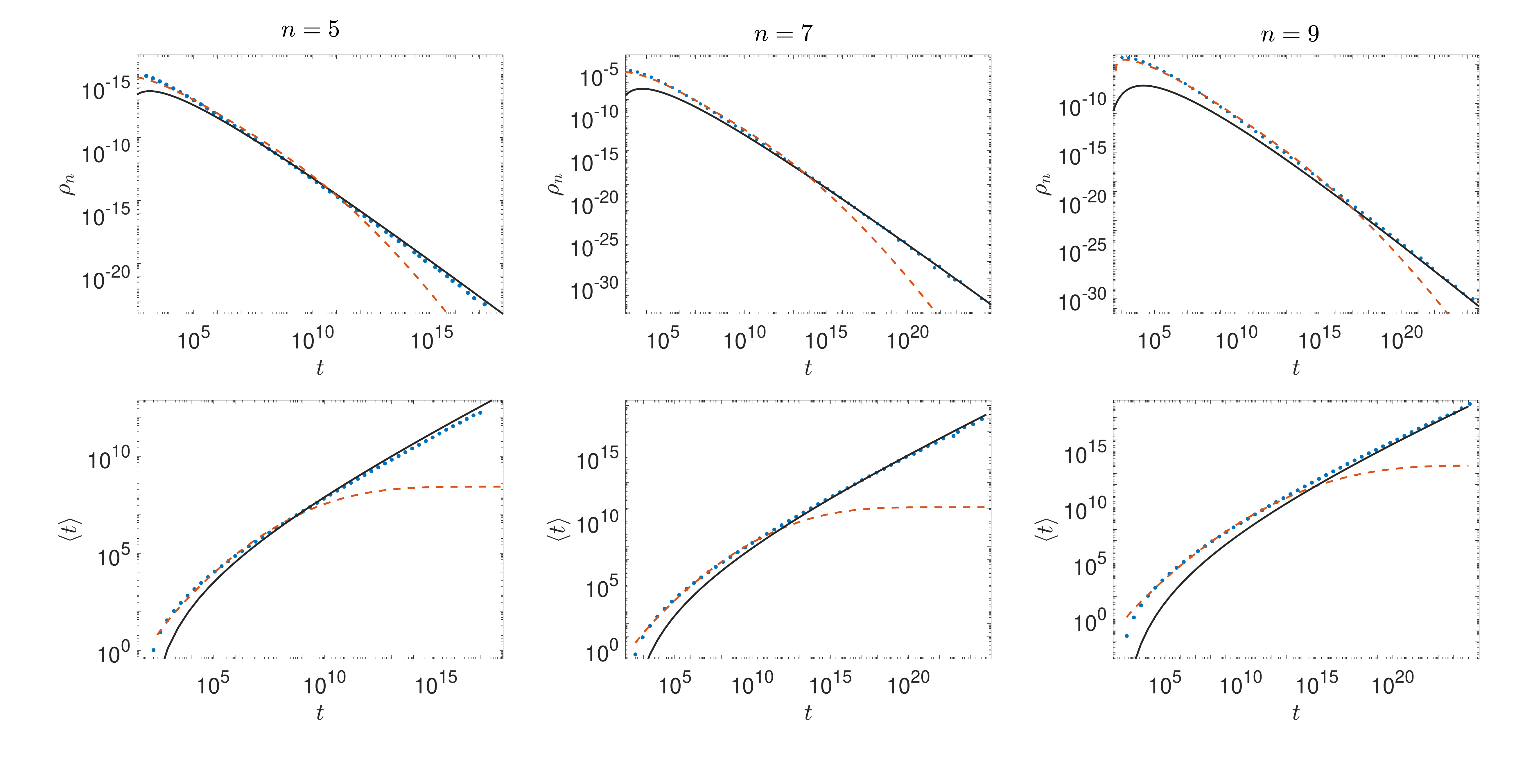}
\caption{\textbf{The influence of rare events in log-aging diffusion.} The probability, \(\rho_n\) are plotted at \(\alpha=0.5\), \(t_0=100\), \(n=5,7,9\) respectively. As \(n\) approaches bigger, the time \(t^\star\) for long decay is bigger and the probability of rare events is less. The dashed red lines are depicted with log-normal distribution \(\frac{1}{t\sqrt{2\pi n\kappa_2}}e^{-\frac{\left(\ln(t/t_0)-n\kappa_1\right)^2}{2n\kappa_2}}\), and the black lines are with long tail \(A\frac{\left(\ln(t/t_0)\right)^{n-1}t_0^\alpha}{t^{1+\alpha}}\), with constant \(A\). But the average \(\langle t\rangle=\int_0^{t}t^\prime\rho_n(t^\prime)dt^\prime\) are dominated by rare events. \(\langle t\rangle_{log-normal}=\int_0^{t}\frac{1}{\sqrt{2\pi n\kappa_2}}e^{-\frac{\left(\ln(t^\prime/t_0)-n\kappa_1\right)^2}{2n\kappa_2}}dt^\prime\) is plotted as red dashed line and \(\langle t\rangle_{log-aging}=\int_0^{t}A\frac{\left(\ln(t^\prime/t_0)\right)^{n-1}t_0^\alpha}{t^{\prime \alpha}}dt^\prime\) is plotted as black solid line.}
\label{supplement_figrho}
\end{figure}

\begin{figure}[b]
\includegraphics[width=1\linewidth]{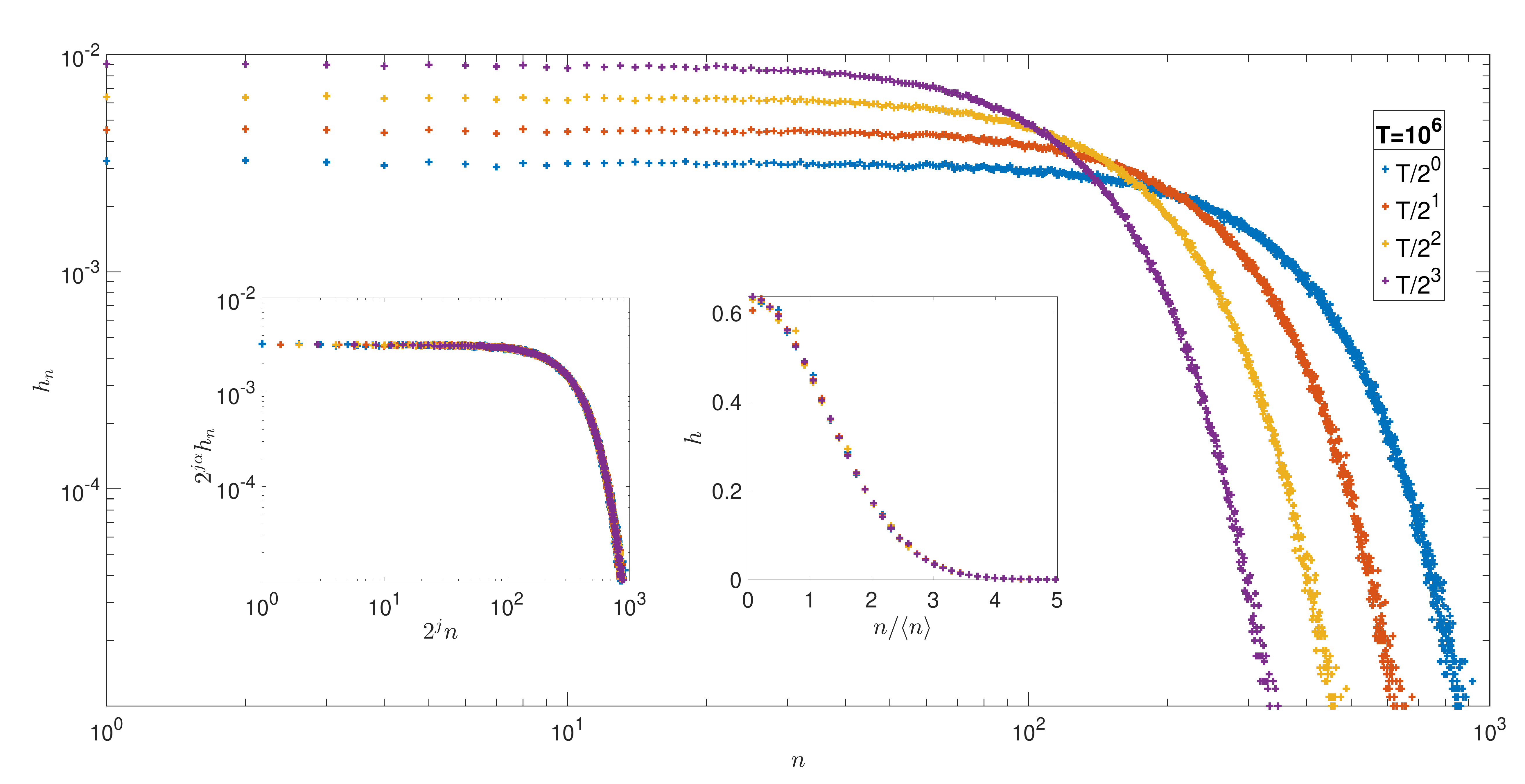}
\caption{Subdiffusion exhibits scaling invariance in the probability \(h_n(t)\) of taking \(n\) steps at different times \(t\). It is shown that the probabilities measured at various times can be rescaled to collapse onto a single universal curve (see the insets).}
\label{supplement_fig.sub}
\end{figure}

\begin{figure}[b]
\includegraphics[width=1\linewidth]{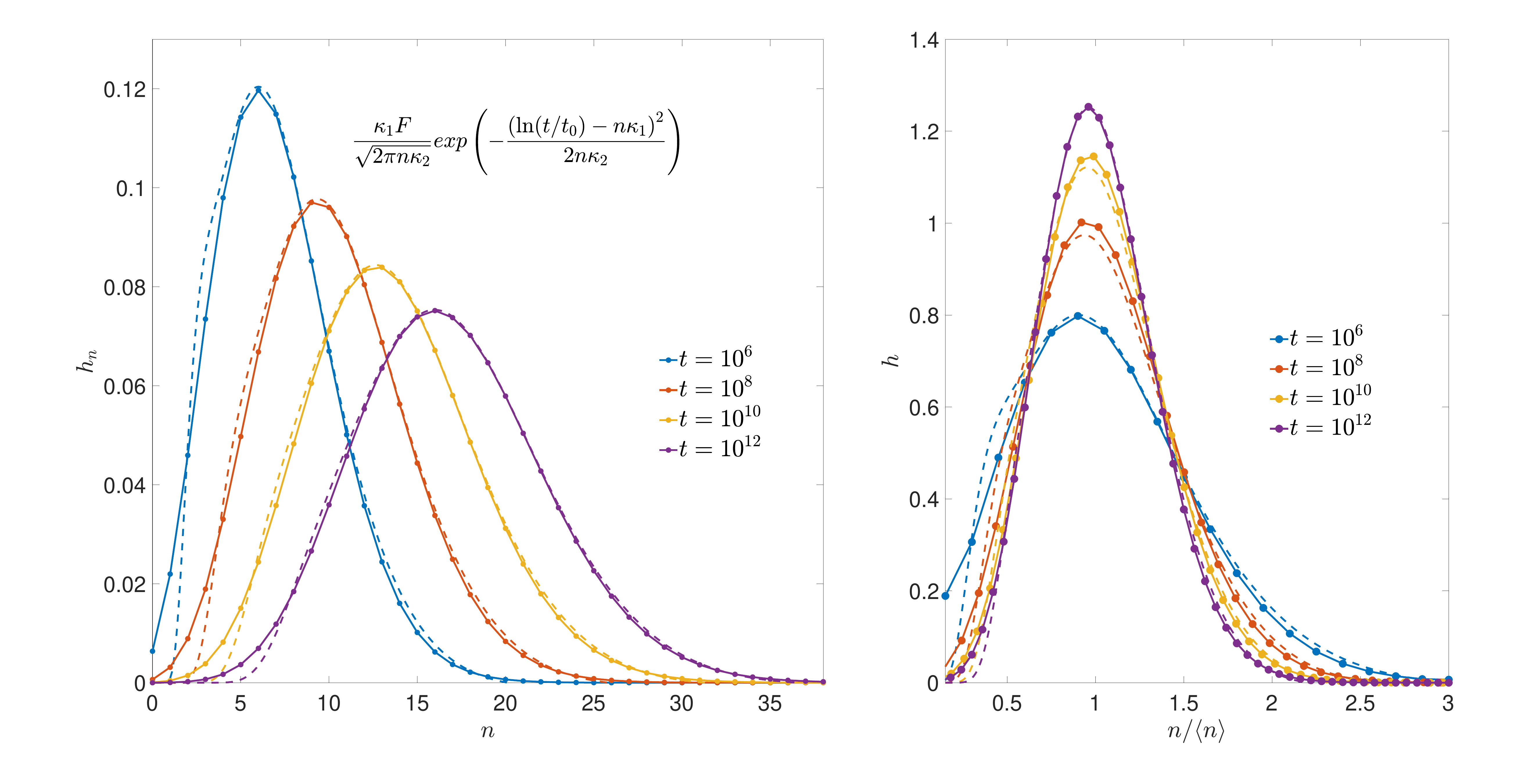}
\caption{(a)The scaling invariance breaking of log-aging diffusion for the probability of jumping \(n\) steps at time \(t\), \(h_n(t)\). (b) The probability, \(h\) versus \(n/\langle n\rangle\). The shape becomes narrow as time goes on, and the fraction beyond the log-normal distribution( dashed line) is less and less.}
\label{supplement_fig.log}
\end{figure}

\begin{figure}[b]
\includegraphics[width=1\linewidth]{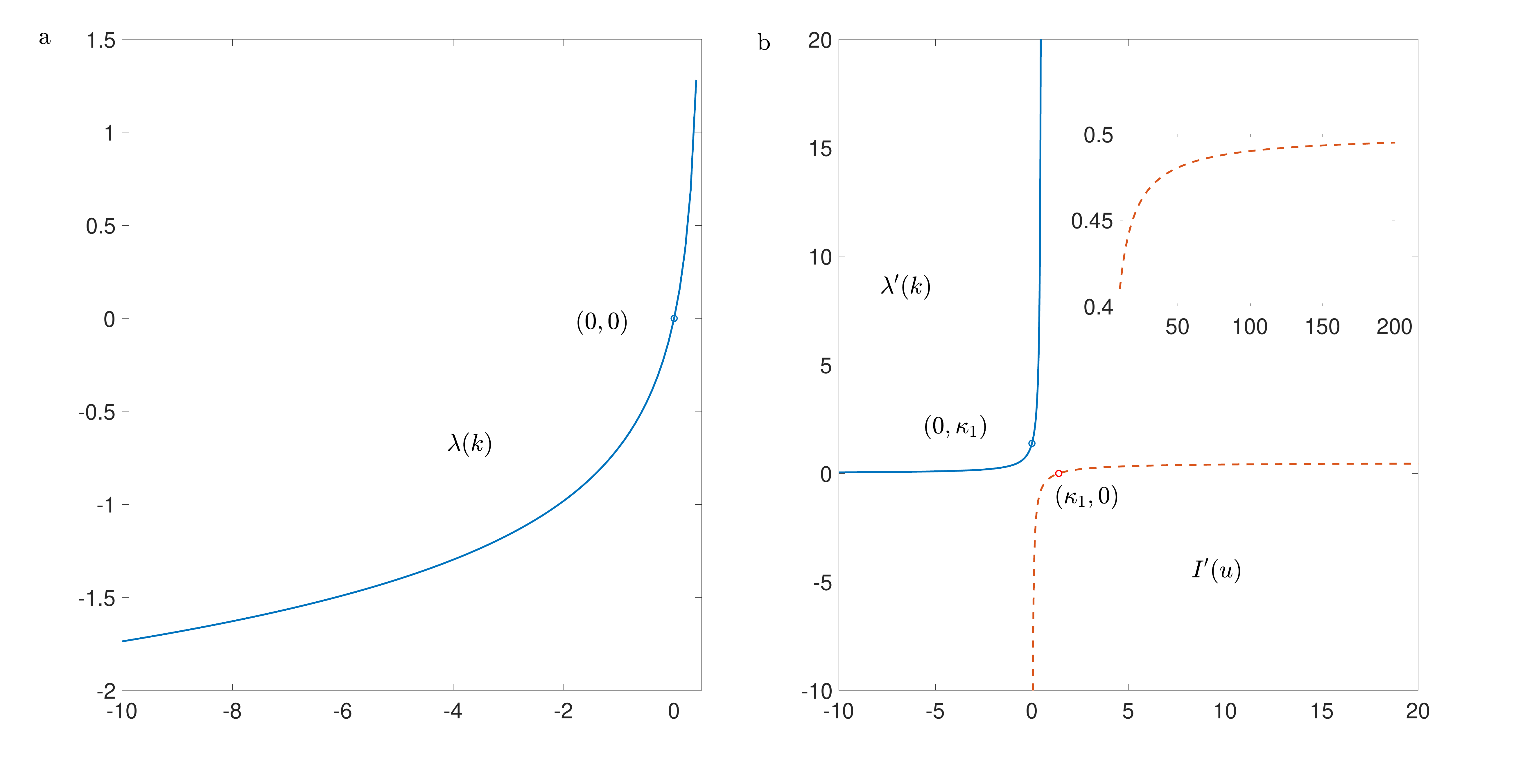}
\caption{The large deviation theory for log-aging diffusion by focusing on the variable \(u=\ln{X}\) with the distribution Eq.~\ref{supple_psiu}. (a) The scaled cumulant generating function \(\lambda(k)\) as a function of \(k\), which exhibits singularities at 
 \(k=\alpha\).  Due to the normalization of the probability, \(\lambda(0)=0\). (b) The deviation, \(\lambda^\prime\); using the Legendre--Fenchel transform (see Eq.~\ref{supple_inversefunction}), one can obtain the corresponding derivative \(I^\prime\) of the rate function \(I\), and the average of the random variable \(u\) can be obtained by \(\langle u\rangle=\lambda^\prime(0)\).  
 }\label{supple_plot}
\end{figure}

\end{document}